\documentstyle[prd,aps]{revtex}
\begin{document}
\draft
\twocolumn[\hsize\textwidth\columnwidth\hsize\csname
@twocolumnfalse\endcsname
\preprint{PACS: 98.80.Cq, FERMILAB--Pub--96/020-A,
IC/97/11\\
{}~hep-ph/yymmnn}
\title{Supersymmetry and Broken Symmetries at High Temperature}
\author{ Antonio
Riotto$^{(1)}$ and Goran Senjanovi\'c$^{(2)}$}
\address{$^{(1)}${\it  Fermilab
National Accelerator Laboratory, Batavia, Illinois~~60510}}
\address{$^{(2)}${\it International Center for Theoretical Physics,
34100 Trieste, Italy }}
\date{February, 1997}
\maketitle
\begin{abstract}
 It is generally believed that   internal symmetries are necessarily 
restored at high temperature in  supersymmetric theories. 
We provide simple and natural counterexamples to this no-go theorem for 
systems having a net background charge. 
We exemplify our findings on  abelian models, for both cases of global
and local symmetries and discuss their possible implications.

\end{abstract}
\vskip1pc]

{\it A. Introduction}. \hspace{0.5cm}    When heated up, physical systems 
undergo phase transitions from 
ordered to less ordered phases. This deep belief, encouraged by 
everyday life experiences, would tell us that at high temperature 
spontaneously broken symmetries of high energy physics get restored.
This, in fact, is what happens in the Standard Model (SM) of electroweak 
interactions. Whether or not true in general is an important question in 
its own right, but it also has a potentially dramatic impact on cosmology.
Namely, most of the extensions of SM tend to suggest the existence of the 
so called topological defects and it is known that two types of such 
defects, {\it i.e.}domain walls and monopoles pose cosmological catastrophe. 
More precisely, they are supposed to be produced during phase 
transitions at high temperature $T$ \cite{kibble} and they simply carry
 too much energy 
density to be in accord with the standard big-bang cosmology.
One possible way out of this problem could be provided by eliminating 
phase transitions if possible. In fact, it has been known for a long time
 \cite{goran} that in theories with a more than one Higgs field (and the
existence of topological defects requires more than one such field in 
realistic theories) symmetries may remain broken at high $T$, and even 
unbroken ones may get broken as the temperature is increased. This offers
a simple way out of the domain wall problem \cite{gia}, whereas the 
situation regarding the monopole problem is somewhat less clear
 \cite{monopole}. Unfortunately, the same mechanism seems to be inoperative 
in supersymmetric theories. Whereas supersymmetry (SUSY) itself gets broken
 at high $T$, 
internal symmetries on the other hand get necessarily restored. This 
has been proven at the level of renormalizable theories \cite{mangano}, and 
a recent attempt to evade it using higher dimensional 
nonrenormalizable operators \cite{tamvakis}, has been shown not to work
 \cite{borut}.

All the papers mentioned above have an important assumption in common: the 
chemical potential is taken to be zero. In other words, they assume 
the vanishing of any conserved charge. On the other hand, it has been known
 that in nonsupersymmetric theories the background charge asymmetry may postpone symmetry restoration 
at high temperature \cite{haberweldon}, and even more remarkably that it can
lead to symmetry breaking of internal symmetries, both in cases of
 global  \cite{scott} 
and local symmetries \cite{linde} at arbitrarily high temperatures. This 
is simply a consequence of the fact that, if the conserved  charge stored
 in the system is larger than a critical value, 
 the charge cannot entirely reside in the thermal excited modes, but it
 must flow into  the 
vacuum.  This is an indication 
that the expectation value of the charged field is non-zero, {\it i.e.} that 
the symmetry is spontaneously broken. 
From the work of Affleck and 
Dine \cite{affleck} we know that there is nothing unnatural about 
large densities in SUSY theories.
The most natural candidate for the large density of the universe is the
lepton number that may reside in the form of neutrinos. In fact, this is
precisley what is assumed in \cite{linde}. Now, one may fear that the
usual washout of $B+L$ number due to sphalerons will  predict the baryon
and lepton numbers to be equal, which would be disastrous. However, there
is a catch here. If the standard model symmetry is not restored at high
temperature, sphaleron effects get exponentially suppressed, and the
usually assumed $B+L$ washout becomes ineffective \cite{ls94}. Thus the
large lepton number can coexist with a small
baryon number.
Furthermore, it should be stressed that a large lepton number is perfectly
consistent with the ideas of grand unification. It can be shown that in
$SO(10)$ one can naturally arrive at a small baryon number and a large
lepton number \cite{hk81}. 

 It is the purpose of this Letter 
to demonstrate that reasonably large charge densities (with the chemical
potential smaller than temperature) provide a natural mechanism 
of breaking internal symmetries in SUSY at arbitrary high temperature.
We exemplify our findings on simple abelian models, with both global
and local symmetries, and leave the generalization to nonabelian 
symmetry for a future publication. However, before presenting our 
results we wish to make some comments regarding background charge asymmetries.

{\it B. Chemical potential: generalities}.\hspace{0.5cm} Let us assume 
that associated with some unbroken abelian symmetry, either
global or local, there exists a nonvanishing net background charge
 $Q = (n - \bar n ) V$, where $V$ is the physical volume of the system
and $n$ and $\bar n$ are the particle and the antiparticle density 
distributions 
\begin{equation}
n = \int {d^3p \over (2\pi)^3} {1\over {\rm e}^{(E-\mu)/T} \pm 1}
\label{enebar}
\end{equation}
and $\bar{n}(\mu)=n(-\mu)$. In the above $\pm$ refers to fermions and bosons respectively and $\mu$ is the 
chemical potential associated with the charge $Q$.
It is easy to show that for fermions (which is all we need here) $
j_0 \equiv \frac{Q}{V} = {g_*\over 6\pi^2}
   \mu^3 (\pi^2 \mu T^2)$ for $\mu\gg T$ ($\mu\ll T$), where $g_*$ denotes the number of degrees of freedom.
Now, if these particles are in equilibrium with the photons, 
then $n - \bar n < n + \bar n \simeq n_\gamma \simeq T^3$ and thus
 $(n - \bar n) / n_\gamma <1$ implies 
$\mu < T$. We shall stick to this reasonable physical assumption.
In general, $\mu$ has temperature dependence dictated by the charge 
conservation constraint. In an expanding Universe $V\sim T^{-3} $ and 
thus the fact that $Q$ remains constant during the expansion implies 
$\mu / T=$ const.  This simple, 
but important observation plays a crucial role in what follows. 

{\it C. Global abelian charge}. \hspace{0.5cm}The simplest supersymmetric
 model with a global $U(1)$ symmetry is provided by a 
chiral superfield $\Phi$ and a superpotential 
\begin{equation}
W = {\lambda\over 3}  \Phi^3.
\label{superpot}
\end{equation}
It has a global $U(1)$ $R$-symmetry under which fields transform as  
$\phi \to {\rm e}^{i\alpha} \phi$ and  $\psi \to 
{\rm e}^{-i\alpha/2} \psi$,
where $\phi$ and $\psi$ are the scalar and the fermionic component of the 
superfield $\Phi$.
Thus the fermionic and bosonic charges are related by 
$2 Q_\psi + Q_\phi = 0$, or 
in other words the chemical potentials satisfy the relation
$\mu\equiv   \mu_\phi = - 2 \mu_\psi$.

Now, the main point here is that the presence of a nonvanishing net
 $R$-charge  leads already at  
{\it tree-level} to a mass term for the scalar field   with a ``wrong''
 (negative) sign after canonical momenta have been integrated out in the
 path integral \cite{haberweldon,scott}. The one-loop high temperature
 effective potential receives dominant contributions  
from $\mu$ and $T$ for a small Yukawa coupling $\lambda\ll 1$ and reads
 (for $\mu<T$) \cite{haberweldon,scott}
\begin{equation}
V_T(\phi) = \left( -\mu^2  +  {1\over 2} \lambda^2  T^2\right) \phi^\dagger
 \phi + \lambda^2 (\phi^\dagger \phi)^2. 
\label{Tpotential}
\end{equation}
The above result is not new; it has been found before 
in the case of charged scalar fields \cite{haberweldon,scott}.
The important new ingredient here is that it holds in the case
of supersymmetry, since
 we may safely  ignore the one-loop fermionic contribution in 
the chemical potential term, being  
 of the order $\lambda^2 \mu^2 \phi^\dagger\phi$  
and suppressed by small $\lambda$. Of course, the fermions enter 
in the $T^2$ mass term and the coefficient 1/2 reflects that.  Obviously, for 
$\mu^2 > \lambda^2 T^2/2$ the symmetry is spontaneously broken at 
high temperature and the field 
$\phi$ gets a vacuum expectation value (VEV)
\begin{equation}
\langle\phi \rangle^2 = {\mu^2 - {\lambda^2\over 2}T^2\over \lambda^2}.
\label{vev}
\end{equation}
This result is valid as long as the chemical potential $\mu$ is smaller 
than the scalar mass in the $\langle\phi\rangle$-background, {\it i.e.}
 $\mu^2 < m_{\phi}^2 = 2 \lambda^2 \langle\phi\rangle^2$ \cite{scott}. 
This in turn implies $\mu > \lambda T$.
In short, for a perfectly reasonable range
\begin{equation}
\lambda T < \mu < T
\label{reasonable}
\end{equation}
the original $U(1)_R$ global symmetry is spontaneously broken 
and this is valid at arbitrarily high temperatures (as long as the 
approximation of $\lambda$ small holds true).  

Notice that, 
in all the above we have assumed unbroken supersymmetry. When supersymmetry
 is softly broken, $U(1)_R$ gets also explicitly broken  because of the
 presence of soft trilinear scalar couplings in the Lagrangian. Therefore,
 the associated net charge vanishes and the reader might be worried 
about the validity of our result. However, the typical rate 
for $U(1)_R$-symmetry 
breaking effects is given by $\Gamma\sim \widetilde{m}^2/T$, where we have 
indicated by $\widetilde{m}\sim 10^2$ GeV the typical soft SUSY 
breaking mass term. Since the expansion rate of the Universe is given
 by $H\sim 30\:T^2/M_{P\ell}$,
$M_{P\ell}$ being the Planck mass,  one finds that 
$U(1)_R$-symmetry breaking effects are in equilibrium and the net charge 
must vanish  only  at  temperatures {\it smaller} than 
   $T_{{\rm SS}}\sim\widetilde{m}^{2/3}M_{P\ell}^{1/3}\sim 10^7$ GeV. 
Therefore, it is perfectly legitimate to consider the presence of a
 nonvanishing $R$-charge at very high temperatures  even in the case of
 softly broken SUSY. 

Thus, we have  provided a simple and natural counterexample to the 
theorem of the restoration of internal symmetries in supersymmetry 
\cite{mangano}. The situation here 
is completely analogous to what happens in the non-supersymmetric scalar field
theory with potential $V=m^2\phi^\dagger\phi+\lambda(\phi^\dagger\phi)^2$ 
described in \cite{scott}. The presence of a nonvanishing chemical potential 
for the  associated $U(1)$ global charge makes it possible for the global 
symmetry to be broken at arbitrarily high temperatures even in the case
 $m^2>0$.   
This is a consequence of the fact that the charge cannot be stored in the 
thermal excited modes, but it must reside in the vacuum and this is an 
indication that the expectation value of the charged field is non-zero, 
{\it i.e.} that the symmetry is spontaneously broken. 

{\it D. Local gauge charge}.\hspace{0.5cm} It has been known for a long time
\cite{linde} that a background charge asymmetry tends to increase symmetry 
breaking in the case of a local gauge symmetry. In his work, Linde has 
shown how a large fermion number density would prevent symmetry restoration
 at high temperature in both 
abelian \cite{linde} and nonabelian theories \cite{linde1}. 
The essential point is that the external charge leads to the condensation
of the gauge field which in turn implies the nonvanishing VEV of the 
Higgs field. This phenomenon may be easily understood if one recalls that 
an increase of an external fermion current ${\bf j}$ leads to symmetry 
restoration in the superconductivity theory \cite{super}. In gauge 
theories symmetry breaking is necessarily a function of $j^2=j_0^2-{\bf j}^2$,
 where $j_0$ is the charge density of fermions. An increase of $j_0$  
is therefore accompanied by an increase of symmetry breaking \cite{linde}. 
We now demonstrate that this phenomenon persists in supersymmetric 
theories, at least in the case of abelian symmetry.

The  simplest model is based on $U(1)$ supersymmetric local gauge symmetry. 
The minimal anomaly free matter content consists of two chiral 
superfields $\Phi^+$ and $\Phi^-$ with opposite gauge charges and the 
 most general renormalizable superpotential takes the form
\begin{equation}
W = m \Phi^+ \Phi^-.
\label{gaugeW}
\end{equation}
Notice that the symmetry is $\it not$ spontaneously broken
at zero temperature.
Since there is no Yukawa interaction, there is also a global 
$U(1)$ $R$-symmetry, under which the bosons have, say, the same charge and
 fermions are invariant. Furthermore at very high temperature, $T> m^{2/3}M_{P\ell}^{1/3}$,
the fermion mass can be neglected and we get also a chiral $U(1)$ symmetry
 under which the bosons are invariant.
We may now suppose for simplicity that there is a net background charge
 density  $j_0$, with the zero current density 
and that it lies entirely in the fermionic sector. In other words we assume
 the background charge to be in the form of the chiral fermionic charge.
Thus only the fermions have a nonvanishing chemical potential.
Equally important, we assume that the gauge charge of the Universe is zero,
just as in \cite{linde}. In the realistic version of this example, one  
would imagine the gauge charge to be the electromagnetic one and the chiral fermionic charge to be, say, the lepton charge in the MSSM. We know from
observation that the electromagnetic charge of the Universe vanishes to 
a good precision. Thus we have to minimize the action with the constraint
that the electric field is zero. What will happen is that some amount of
 bosonic charge will get stored into the vacuum in order to compensate 
for the fermionic one and achieve the vanishing of the electric field.
In this way, the total $U(1)$ charge density of the system including 
the charge of the condensate is equal to zero even if symmetry is broken and
 the gauge forces are short-range ones. 
This is the principal reason behind the resulting
spontaneous symmetry breaking 
of the local gauge symmetry, as we show below. It is crucial thus to have 
some nonvanishing external background charge, {\it i.e.} the model should have 
some extra global symmetry as provided by our chiral symmetry.  
We can obviously take $A_i=0$ in the vacuum and treat $A_0$ on 
the same footing with the scalar fields $\phi^\pm$ (due to the net charge 
$A_0$ cannot vanish in the vacuum). If we now  integrate out $A_0$ using its
equation of motion, assuming the electric field to be zero, we can then 
 compute the high temperature potential for the scalar fields in question 
at high temperature and large charge density  with the following result
 \begin{eqnarray}
V_{\rm eff}(T) &= &{g^2 \over 2} T^2\left(|\phi^+|^2 + |\phi^-|^2\right) 
+{g^2 \over 2}  \left(|\phi^+|^2 - |\phi^-|^2\right)^2 \nonumber \\
&+ &  {1 \over 2} {j_0^2 \over 2 \left(|\phi^+|^2 + |\phi^-|^2\right) + T^2},
\label{eff}
\end{eqnarray}
where we have taken $T\gg m$ and we have included both scalar and fermionic
 loop contributions in the $T^2$ mass term for $A_0$. 
Now, except for the $D$-term, the rest of the potential depends only on the 
sum $\phi^2 \equiv |\phi^+|^2 + |\phi^-|^2 $, and thus the energy is 
minimized for the vanishing of the $D$-term potential, {\it i.e.} for 
$|\phi^+|^2 = |\phi^-|^2$. It is easy to see that in this case the 
effective potential has two extrema: 
\begin{equation}
\phi = 0 \quad {\rm and} \quad \phi^2 = {j_0 \over\sqrt{2} g T} 
- {T^2 \over 2}.
\label{vevT}
\end{equation}
The second extremum obviously exists only for 
\begin{equation}
j_0 > {g T^3\over\sqrt{2}}.
\end{equation}
Moreover in that case it is an absolute minimum, while $\phi = 0$ is
a maximum.  
% (see below for the opposite case of the scalar charge) and  
%we decompose the scalar  fields as
% $\phi^{\pm}(x)= 1/\sqrt{2}\left[\phi_1^\pm(x)+i\:
%\phi_2^\pm(x)+\langle \phi_1\rangle^{\pm}\right]$ and the gauge 
%field as $A_\mu(x)=B_\mu(x)+C_\mu$, $\langle\phi_i\rangle^{\pm}$ and
% $C_\mu$ being  
% the vacuum expectation values in the thermal ensemble. The situation 
%here is analogous to the one of Linde \cite{linde}; 
%however with the important difference that our model is supersymmetric.
%
%
%The equation of motions are given by 
%\begin{eqnarray}
%\sigma_{\pm}\left[-m^2(T)+g^2 \: C_0^2-\frac{g^2}{2}
%\left(\sigma_{+}^2-\sigma_{-}^2\right)\right]&=& 0,\\
%g \: C^0\left(\sigma^2+\frac{1}{3}T^2\right)-J^0&=&0,
%\label{system}
%\end{eqnarray}
%where $m^2(T)=m^2+g^2T^2/3$ and $\sigma^2=\sigma_{+}^2+\sigma_{-}^2$.

Now,  we can rephrase the above condition in the 
language of the chemical potential (using  $g_* = 4$ ): $
\mu > g T$.
For $ g\ll1$, which is of our interest, $\mu $ easily satisfies 
the condition $\mu < T$. As noted above, since Yukawa interactions are
 absent, the role of the external charge may have been played  by the
 $R$-charge in the scalar sector, the two scalars being equally charged
 under this symmetry. In such a case, the analysis requires  careful 
handling because of issues related to gauge invariance and 
we will extensively  explore this in a future publication.  We only 
mention here that the addition of the term
$-\mu\left(\phi_{+}^{\dagger}\stackrel{\leftrightarrow}{\partial}_0\phi_{+}+ 
\phi_{-}^{\dagger}\stackrel{\leftrightarrow}{\partial}_0\phi_{-}\right)$ to the
 Lagrangian   requires a simultaneous addition of the term $2 
A_0\mu\left(|\phi_{+}|^2-|\phi_{-}|^2\right)$ to conserve gauge invariance. 
It is straightforward to show that, for large enough chemical potential $\mu$,
 the effective potential at high temperature is again minimized for vanishing 
$D$-term, which results  into $A_0=0$, and that $|\phi|^4$-terms induced at 
one-loop are crucial for 
the existence of global minimum which breaks local gauge symmetry. For the 
physical realization of this situation the existence of an extra source is
 necessary and, in our case, the latter is provided by the chiral fermionic
charge.
This
 situation is quite different  from what happens in a simple nonsupersymmetric
 model with local abelian symmetry and only one scalar field, where only the 
combination $(A_0+\mu)$ appears in the computation and  the absence of an 
extra source imposes the condition that this quantity is equal to zero 
\cite{lindek}. 

Thus we have also provided a supersymmetric example of a local gauge 
symmetry being broken at high temperature in the presence of a background 
charge density. 

{\it E. Summary and outlook}. \hspace{0.5cm}  The main point of our paper
is that internal symmetries in supersymmetric theories,
 contrary to the general belief, may be broken at high temperature, as
long as the system has a nonvanishing background charge. The examples 
we have provided here, based on both global and local abelian symmetries,
are natural and simple and should be viewed as prototypes of more
realistic theories. The necessary requirement for the phenomenon
to take place is that the chemical potential be bigger than a fraction
of temperature on the order of (1-10)\%. Notice that this is by no 
means unnatural. In the expanding universe, as we have stressed before,
the chemical potential is proportional to temperature, and thus unless
zero for some reason, $\mu /T$ is naturally expected to be of order
one. More important, this chemical potential could be zero today, all
that is needed is that it is nonvanishing at high temperature. We 
have seen how soft supersymmetry breaking may naturally provide such a 
scenario if there is some nonvanishing external charge.

 Now, it is well known that in suspersymmetry the existence of flat 
directions may lead to large baryon and lepton number densities at very
high temperature \cite{affleck}. In any case, we wish to be even more
open minded and simply allow for a charge density without worrying about 
its origin. It is noteworthy that a large neutrino number density
may persist all the way through nucleosynthesis up to today 
\cite{chemicalpotential}. This 
has been used by Linde  \cite{linde1} in order to argue that even in SM
the $SU(2)_L\otimes U(1)_Y$ symmetry may not be restored at high temperature. 
Since SUSY, as we have seen, does not
spoil the possibility of large chemical potentials allowing symmetry 
breaking at high $T$, it is important to see if our results remain 
valid in the case of nonabelian global and local symmetries. This work
 requires particular attention due to the issues related to gauge invariance
 and is now in progress. We only wish to observe here that, if 
some conserved charge is present in the system, {\it e.g.} the lepton
 charge in the minimal supersymmetric extension of the SM,  it will be 
automatically shared among fermions and scalars of the same 
supersymmetric multiplet. However, in the realistic situation in which 
supersymmetry  is softly broken, sfermions acquire a 
mass $\widetilde{m}$ and are much heavier than their 
fermionic partners.  Hence, for temperatures
$T$ smaller than $\widetilde{m}$, the number density of 
sfermions in the thermal bath is drastically reduced by the 
Boltzmann suppression factor and the external conserved charge 
will reside in the fermionic sector. This might render the  analysis easier 
and make it  similar to the one performed in Section {\it D}.    

We should stress that there is more than a sole academic interest
to the issue discussed in this paper. If symmetries remain broken at 
high temperature,
there may be no domain wall and monopole problems at all. Furthermore,
it is well known that in SUSY  grand unified theories symmetry restoration
 at high temperature prevents the system from  leaving the false vacuum and 
finding itself  in the broken phase at low $T$. This is a direct consequence 
of the vacuum degeneracy characteristic of supersymmetry which says that at
 zero temperature the SM vacuum and the unbroken GUT symmetry one have the same
(zero) energy.
If the symmetry is restored at high $T$, one would start with the unbroken
 symmetry in the early universe and would thus get caught in this state 
forever. Obviously, if our ideas 
hold true in realistic grand unified theories, this problem would not
arise in the first place.

    To conclude, our abelian examples, based on both global and local
gauge symmetries, show that it is perfectly consistent to have a spontaneously 
broken internal symmetry in supersymmetric theories. For this to happen,
 though, an important condition must be satisfied: the system must possess
 a non negligible chemical potential, $\mu \leq T$, or in other words a net 
background charge. 
This is the novel and the main point of our paper. The previous works,
 it should be stressed, indicating the role of chemical potential in 
symmetry nonrestoration at high temperature  referred only to 
nonsupersymmetric theories. Now, in nonsupersymmetric theories, in many 
cases it is possible to achieve high temperature symmetry breaking even
 without resorting to charge asymmetries, whereas it is impossible in SUSY. 
 Generalization of our results to supersymmetric nonabelian theories is now
 in progress. 

\vskip 0.3 cm

We are deeply grateful to Borut Bajc for his keen interest in our work,
for many useful discussions, for checking some of our computations and for 
the careful reading of the manuscript. We also wish to acknowledge discussions
 with  Graciela Gelmini, Andrei Linde, Alejandra Melfo  and, in particular, 
Scott Dodeldson.   AR is supported by the DOE and NASA under Grant NAG5--2788.


\begin{references}
\bibitem{kibble}T.W.Kibble, {\it J.Phys.}, {\bf A9} (1976) 1987;
               {\it Phys. Rep.} {\bf 67} (1980) 183.
\bibitem{goran}S.Weinberg, {\it Phys. Rev.}, {\bf D9} (1974) 3357;
                 R.N.Mohapatra and G.Senjanovi{\'c}, {\it Phys. Rev. Lett.}
              {\bf 42} (1979) 1651; {\it Phys. Rev.} {\bf D20} (1979) 3390.
\bibitem{gia}G. Dvali and G. Senjanovi\'c, {\em Phys. Rev. Lett.}
               {\bf 74} (1995) 5178.
\bibitem{monopole}G. Dvali, A. Melfo and G. Senjanovi\'c, {\em Phys. Rev.
 Lett.}   {\bf 75} (1995) 4559; G. Bimonte and L. Lozano, {\em Phys.Lett.}
 {\bf B366},248   (1996); {\em Nucl.Phys.} {\bf B460}, 155 (1996).
\bibitem{mangano} H. Haber, {\it Phys. Rev. } {\bf D26} (1982) 1317 ;  M. 
              Mangano {\it Phys.Lett.} {\bf B147} (1984) 307
\bibitem{tamvakis} G. Dvali and K. Tamvakis, {\it Phys. Lett. } {\bf B378}
              (1996) 141.
\bibitem{borut}B. Bajc, A. Melfo and G. Senjanovi\'c, {\it Phys. Lett. }
 {\bf B387} 
             (1996) 796, hep-ph/9607242.
\bibitem{haberweldon} H.E. Haber and H.A. Weldon, {\em Phys. Rev.} {\bf D25}
 (1982) 502. 
\bibitem{scott} K.M. Benson, J. Bernstein and S. Dodelson, {\em Phys. Rev.} 
{\bf D44} (1991) 2480. 
\bibitem{linde} A.D. Linde, {\em Phys. Rev.} {\bf D14} (1976) 3345.

\bibitem{affleck} I. Affleck and M. Dine, {\em Nucl. Phys.} {\bf B249} (1985)
361.

\bibitem{ls94} J. Liu and G. Segr\`e, {\em Phys. Lett. } {\bf B338}
(1994),  259.

\bibitem{hk81} J. Harvey and E. Kolb, {\em Phys. Rev} {\bf D24} (1981),
2090.

\bibitem{linde1} A.D. Linde, {\it  Phys. Lett.} {\bf B86} (1979) 39.


\bibitem{super} See for instance, P.G. De Gennes, {\em Superconductivity
 of Metals and Alloys} (Benjamin, New York, 1966).
\bibitem{chemicalpotential} Large neutrino chemical potentials $|\mu|$ for
 the three
 neutrino species affects the energy density in the neutrino species, 
and thereby the expansion rate of the Universe, dangerously 
increasing the primordial production of $^4$He. However, it is possible to 
dial in the desired $^4$He abundance by the appropriate sign of the neutrino
 chemical potential, see, for instance E.W. Kolb and M.S. Turner, 
{\em The Early Universe}, Addison-Wesley Ed. (1990), p. 107. 
\bibitem{lindek} D.A.  Kirzhnits and A.D. Linde, {\em Ann. of Phys.} {\bf 101} 
(1976) 195. 
\end{references}
\end{document}